# The myths of the Bear[1]

Elio Antonello
*INAF – Osservatorio Astronomico di Brera*
*SIA – Società Italiana di Archeoastronomia*
elio.antonello@brera.inaf.it

**Abstract.** Following previous works on ancient myths in Greek and Latin literature regarding Ursa Major, and the possible relation with the ancient shape of the constellation, we discuss further this case in the light of the evolution of *Homo sapiens* and the ethnographic records of populations of Eurasia and North America.

## 1. Introduction

Since the remotest antiquity, our ancestor *Homo sapiens* might have been impressed by the spectacular sight of the night sky. He had a 'physical' knowledge of the surrounding world except for the sky: he could observe the Sun and the other celestial bodies with their movements, but those objects were out of reach. That should have been a strong stimulus to the human intellect, forced to find explanatory myths of what he was observing. During the clear nights, especially when the Moon was not visible, the splendour of myriads of stars should have stirred his imagination. Today it is difficult to appreciate such a spectacular night sky because of the light pollution, but it had been possible to enjoy fully it until two hundred years ago. Presently, we have to go to the most remote places and deserts, hundreds of kilometres away from towns, to try at least part of the deep emotion probably felt by our ancestors. They should have associated asterisms and constellations with the shapes of real or fantastic animals, and that would be attested by the myths and the legends handed down by many populations.

The group of seven bright stars of Ursa Major, i.e. the Great Bear (she-Bear), is a common cultural heritage of various populations of Eurasia and North America. On the basis of the Greek myth of the nymph Callisto, we tried to illustrate how the identification of this group with a bear could have been related to the shape of the constellation some tens of thousand of years ago (e.g. Antonello 2008, 2009)[2]. In the present note we summarize the previous works and discuss some ethnographic records that are relevant to this issue.

## 2. The constellations of the Bears

Figure 1 shows a simulation of the sky as it can be seen today, with Ursa Major at the centre, the bright star Arcturus in the constellation of Boötes, and Ursa Minor (Little she-Bear) with the Polar Star. The name Boötes indicates a cowherd, and this makes sense if one considers the other name of the constellation Ursa Major, i.e. Big Wagon or Chariot: indeed, it could be the oxcart led by Boötes. However, we are interested in the presumably older name of the constellations, when chariots had not yet been invented, nor there was the breeding of cattle and even the agriculture. Our ancestors could have identified the constellation as a bear during the Stone Age, at the time of hunters-gatherers, because it is a common representation to several populations of Europe, Asia and to natives of North America; that is, the identification might date back to the Ice Age, when Eurasia and America formed a single continent (Gingerich 1984).

The Greek term *arktos* means "bear"; well-known words were derived from it, such as Arctic, to indicate the land under the sky of the Bears, and also the name of the star Arcturus, which means bear guard[3]. However, this last name may appear unjustified, because Arcturus is not right next to the bears.

---
[1] Based on talks given at the archaeoastronomy meetings "La misura del tempo", University of Sassari (Sardinia), 12 december 2011, and "Cielo e cultura materiale. Recenti scoperte di archeoastronomia", XIII Borsa Mediterranea del Turismo Archeologico, Paestum, 20 november 2010.
[2] The astronomical reconstruction raised some interest among archaeologists and it was included in the catalog of an archaeological exhibition (Antonello 2008).

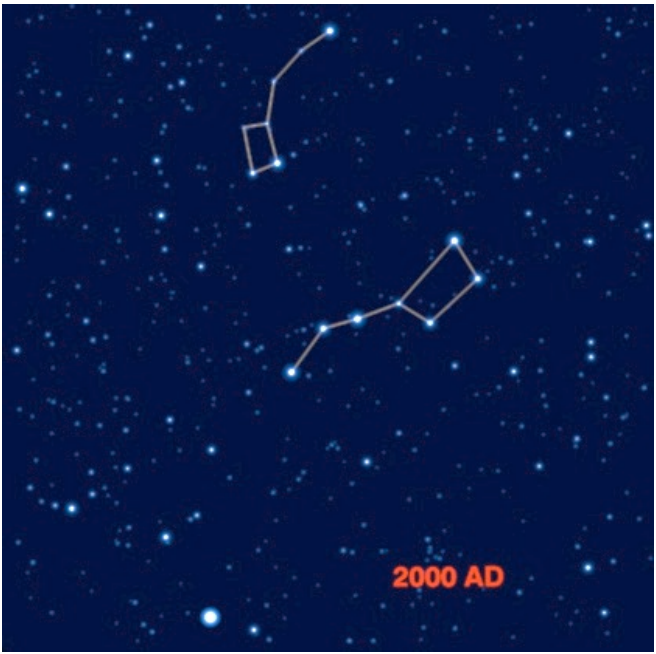

**Fig. 1.** The constellations Ursa Minor (top) and Ursa Major (centre); the brightest star in the lower part of the field is Arcturus in the constellation Boötes.

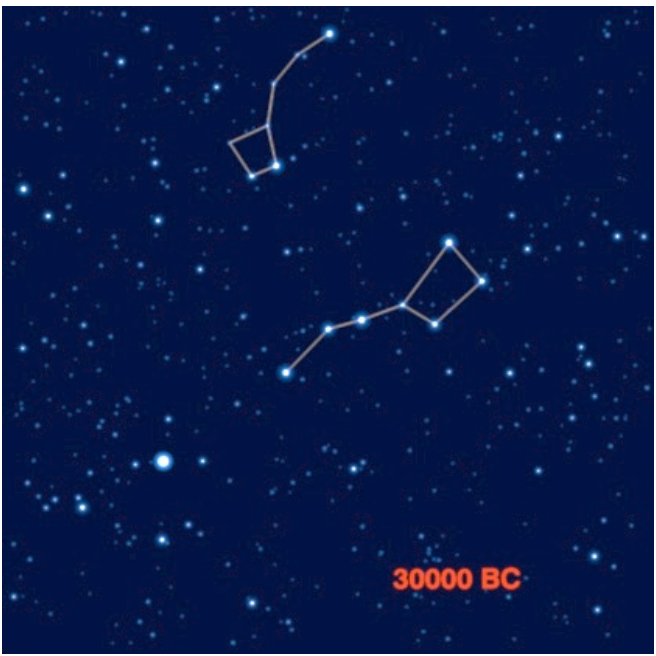

**Fig. 2.** The Ursa Minor, Ursa Major and Arcturus as they were visible around 32,000 years ago (to be compared with Figure 1).

The constellations change their shape, albeit very slowly, over the course of tens of thousand of years because the stars have their own proper motions. Each star moves with its own velocity in the Galaxy, and because its distance from Earth is generally very large, the apparent motions are tiny, and very accurate astronomical observations are needed to measure them. Using the parallaxes and proper motions of stars obtained with observations from the space satellite HIPPARCOS, and accurate determinations of stellar radial velocities, one can go back in time and reconstruct the shapes of the constellations as they were seen in the distant past by our ancestors (for some technical details, see Antonello 2010a,b). For example, Figure 2 shows the appearance of this area of the sky approximately 32,000 years ago. From a comparison with Figure 1 one can see that Arcturus was then located closer to the Bears; that is, the name of this star appears to be better justified if it is referred to the very far past. To go back so much, though, means to get in environmental and climatic conditions totally different from those found in historic times, and it is likely that such different conditions have had some effect on the formation of specific myths and legends.

---

[3] According to Blomberg (2007), however, in ancient times the term *arktos* may have indicated just the northern direction, since the identification with a bear appears to be reported only by later writers.

### 3. The climate from Palaeolithic to Neolithic

According to most anthropologists our species *Homo sapiens* first appeared in Africa many tens of thousand of years ago (probably more than 100,000 years ago), and he would have had our physical characteristics; in particular, it may be reasonable to assume that he have had our intellectual faculties[4]. Obviously he had not all the knowledge that will be accumulated over the course of tens of thousand of years[5]. Our ancestors probably left Africa about 50,000 years ago, and spread across Europe, where the Neanderthals were living, between 47,000 and 41,000 years ago (Mellars, 2006). They were able to create wonderful paintings such as those found in the cave of Lascaux in France (about 16,000 BC), and to produce female (and other) figurines, such as the bas-relief of Lausselle in France (approximately 22,000 BC), and the Venus found in Moravia (Dolní Věstonice, 27,000 BC)[6]. Among the even older artistic expressions there are the paintings of the cave Chauvet-Pont-d'Arc (France), dated to 30,000 BC. During the Palaeolithic, at our latitudes, our ancestors grouped together in small societies (nomadic tribes), and subsisted by gathering plants and hunting or scavenging wild animals, to eat their meat and to cover themselves with their skins[7]. Of course, also the bears were among their preys, and we will try to give support to the view that at least some myths of the Bears may have arisen during the Palaeolithic age, and have survived until historic times in spite of the economic and social revolutions occurred during the Neolithic.

The first known Neolithic settlements, with the earliest forms of farming, have been found in the Middle East and date back to about 11000 years BC. If our ancestors were similar to us, why did that really impressive change occur only tens of thousand of years after the migration out of Africa? Renfrew (2007) raised that question as the problem of the 'Sapient Paradox'. The main reason could be probably attributed to the climate changes, that forced the development of a different economy based on agriculture and farming after the Ice Age (Cauvin 2000)[8]. In order to assess the hypothesis of the formation of the myths of the celestial Bear before the thaw it is important to understand the origin of these climate changes and their effects.

Detailed analysis of the drillings of oceanic floors, of ice in Antarctica and Greenland, of lake floors and other sediments carried out over the past forty years has provided consistent and coherent information on the climate of the past several hundred thousand years. The proxies are essentially local indicators, but on a long enough timescale they should reflect the average climatic conditions on Earth. The diagrams in Figure 3 show the variation of some climatic indicators during the past 40,000 years. It is shown the change of temperature in Antarctica, derived from the analysis of ice sheets (deuterium isotope; Petit et al. 1999), the average temperatures for the Congo River basin (analysis of distribution of lipid of tetraetere bacteria; Weijers et al. 2007), and the temperature of Greenland (isotopes in ice; Alley 2004). One can see how the average temperature of the Earth changed relatively little over the last 12,000 years, while there was an evident change between 16,000 and 14,000 years ago, when it increased after a long cold phase. The variations are more significant going from the equatorial zone to the poles.

---

[4] "Early *H. sapiens* were as cognitively advanced as those today. Differences in the most ancient artifacts did not reflect a different level of cognition in their makers, but simply the need to create objects to suit different environmental and social conditions" (Shea, in Balter 2013).
[5] This process should have implied also a loss of other knowledges. For instance, an "average individual in a hunting or fishing culture had much greater practical knowledge about the animal world" than those in most modern societies (Darnell 1977, quoted by McLaren et al. 2005).
[6] It may be worth to remark that the Venus of Dolní Věstonice was realized using a baked clay technique.
[7] The diet of people living near rivers, lake and sea shores included fish and shellfish.
[8] Watkins (2010) pointed out that probably the sedentism begun at least about 25,000 years ago (Early Epipaleolithic) in Near East; therefore farming should have complemented rather than prompted the advent of permanent communities.

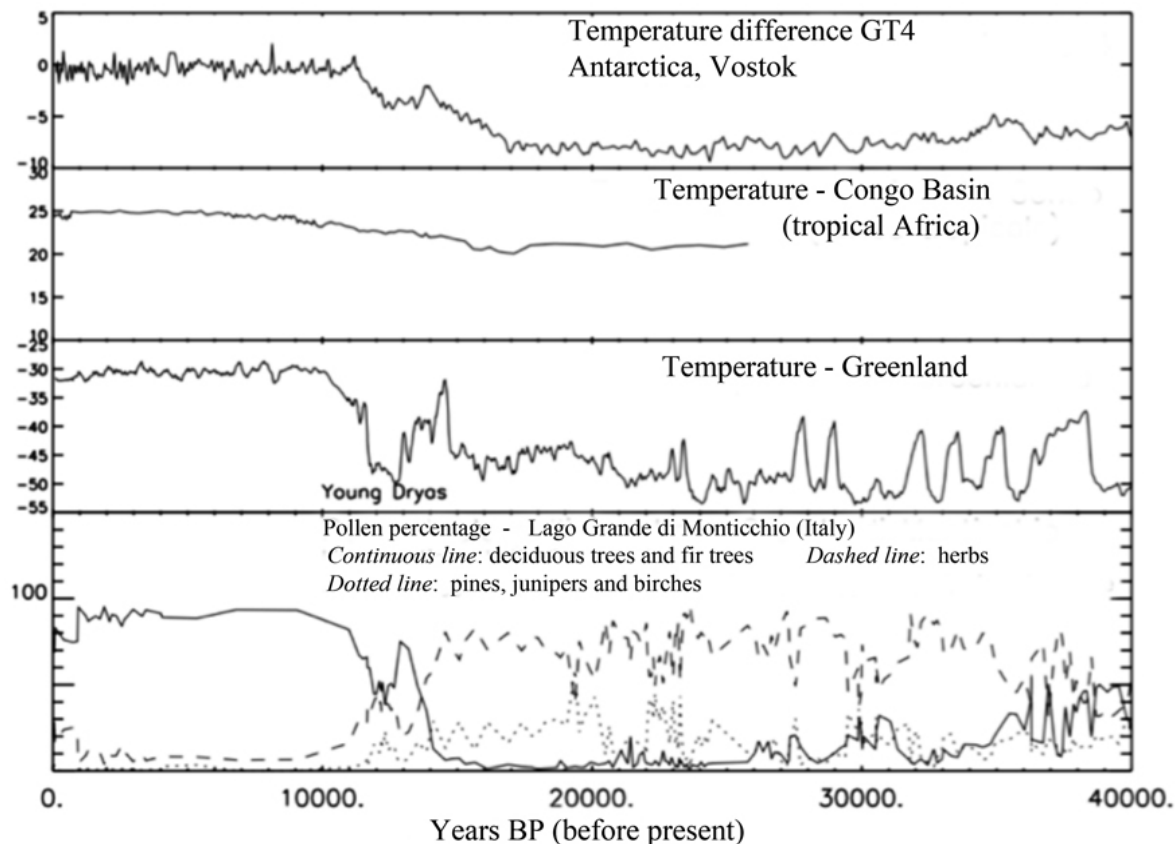

**Fig. 3.** Average temperature trends in Antarctica, tropical Africa, and Greenland, and the percentage of pollen in Italy in the last 40,000 years. Usually the BP year 0 (left) is 1950. One can see how the temperature increased starting about 16,000 years ago, with a corresponding complete change of vegetation. In the northern hemisphere there was a cold period (Young Dryas) about 12,000 years ago, which lasted a thousand years. *Antarctica*: data by Petit et al. (1999); *Congo basin*: data from Weijers et al. (2007); *Greenland*: data from Alley (2004); *pollen data* from Allen et al. (1999). See the cited works for a discussion of the time scale and temperature calibrations considered.

The stratigraphic analysis of pollen deposited in the Lago Grande of Monticchio, located in the crater of Monte Vulture in Basilicata (Italy), provides another extremely interesting indication (Allen et al. 1999). The graph shows the percentage of pollen of various plants. The pollen of deciduous trees (such as oaks) and fir trees is the most abundant one during the last 12,000 years, while it was much less in the previous epoch[9]. There were much more herbaceous plants than trees during the Palaeolithic epoch, when the presence of mountain (or cold climate) trees, such as pines, junipers and birches, was also more significant. From these diagrams one can understand the enormous change occurred over the few millennia around 12,000 years ago. Before that date the weather was cold, dry and unstable: pollens indicate few trees, mostly of mountains or cold climates, and a lot of grass, while in the last 11,000 years the Vulture was characterized mainly by forests of temperate climates. During the Palaeolithic the local environment should have been characterized by steppe, which is typical of present day alpine or Siberian tundra.

---

[9] The fairly uniform trend of the pollen percentage from 10,000 years ago until historic times (with recent changes that may be probably of anthropic origin) is confirmed by the more detailed data by Allen et al. (2002).

Since for many thousand years the climate was cold, dry and unstable, it is plausible to think that something similar to farming could not even be imagined, because the crops need warm moist seasons that must be fairly stable, i.e. just the climate of the last 10,000 years or so[10]. We remark that the palaeoclimatologists generally agree on the dependence of such a climate change on the orbital forcing. In other words, the long-term periodic behaviour of the terrestrial climate depends on the astronomical parameters of the Earth orbit (at least on a time interval of some million years). It is impressive that apparently there are no reasonable alternative theories to this one. Therefore, one should conclude that the astronomical effects triggered the evolution of the human society.

During the Palaeolithic the sea level was lower by more than about 120 m than the present value, since there was ice instead of water[11]. For example, during this period, Italy was quite different: the Po Valley extended to half of present day Adriatic Sea, the Elbe island was joined to Tuscany, Corsica to Sardinia, and Sicily to Egadi and to Maltese Islands. Asia and North America were a single land, and, during the thaw, the formation of the Bering Strait between Siberia and Alaska separated the two continents. Before that, however, *Homo sapiens* could have gone back and forth in Asia and North America, possibly following the large herds. As suggested by Gingerich (1984) the identification of the constellation with a celestial Bear, a common culture to the natives of Eurasia and North America, therefore could have occurred before the thaw. That is, such identification could date back to the Palaeolithic epoch (see also Schaefer 2006).

**4. The relation between man and bear**

In the last century the paleoanthropologists attempted to characterize the possible Palaeolithic lifestyle by borrowing the descriptions of indigenous populations of modern epoch from the ethnographic studies[12]. It may be possible that also in the far past the bear could have been a totemic animal as in the modern times, that is, the animal may have been considered both an important hunting prey and a deeply respected being. Many of the native tribes in North America and Eurasia had with the bear a rather different relationship than with other creatures, and the reports of explorers, travellers and ethnographers all agree on the peculiar respect enjoyed by such an animal (Hallowell 1926). There is an extensive literature on the unusual characteristics of the relationship with the bear, such as the fact that he was considered an ancestor, or a distant relative, with a superior intelligence and smarter than his "descendant", the man. He was often considered a kind of relative (grandfather or grandmother), or the oldest person of the tribe. As summarized by Germonpré & Hämäläinen (2007), people have felt a kind of kinship with bears since humans and bears share many characteristics. They live in the same regions and eat the same fish, roots, and berries. Unlike other animals, bears can stand on their hind legs as humans do and they can use their fore paws as humans use their hands. A bear's skinned body looks human, and several bear bones resemble human bones, which lends credence to the view that the animal is really a man in disguise, while transformations between humans and bears can be found in many tales of North America tribes. The skin of the bear is the bear's winter fur coat and it can be taken off in the same manner as people's clothes. In Eurasia, people believed that the bear was the only animal that has a human-like soul.

---

[10] At least this should have been the case for the lands not located in the tropical/equatorial zone. Recent data from the New Guinea Highlands demonstrate the exploitation of the endemic pandanus and yams in archaeological sites more than 40,000 years ago; the early inhabitants probably cleared forest patches to promote the growth of useful plants (Summerhayes et al. 2010). In this case, however, *Homo sapiens* exploited the wild (not domesticated) tubers and plants, that is, it was not yet agriculture.

[11] One should take into account also global geophysical effects such as the pressure of ice masses on the magma distribution under the Earth crust.

[12] However, the peoples called "primitive" in the ethnographies are not "primitive" in the true meaning of the word, since each of the elements of their culture was the result of a long process of development (Anisimov 1963 p. 159). Also Renfrew remarked that it is unreliable to generalise from the ethnographic present to the palaeolithic past without explicit support evidence. "Modern hunter-gatherers have had as long as any other contemporary communities to develop from our common palaeolithic predecessors, and their culture is as distant in time from the life and times of the palaeolithic as is ours" (Renfrew 2007, p. 141).

The Ainu or Aino of Japan are an indigenous population that maintained animistic beliefs until recent times, and some anthropologists think that the bear worship during Palaeolithic should have been similar to that of the Aino. Frazer (1922), however, pointed out that the bear can hardly be described as a sacred animal of the Aino, nor yet as a totem; for they do not call themselves bears, and they kill and eat the animal freely. On the other hand, they have a legend of a woman who had a son by a bear; and many of them who dwell in the mountains pride themselves on being descended from a bear: "As for me, I am a child of the god of the mountains; I am descended from the divine one who rules in the mountains," meaning by "the god of the mountains" no other than the bear. It is therefore possible that the bear may have been at least the totem of an Aino clan (Frazer, 1922, p. 505-506)[13].

Other anthropologists have suggested that the tribal structure in the Palaeolithic epoch was matriarchal[14]; hence, taking into account the possible origin of ancestral ceremonies regarding the bear, one might conclude that it was probably the she-bear (great mother) and not the male bear to be represented in the sky, tens of thousand of years ago (Gurshtein, 1993; see also the references therein regarding the works of Russian authors on Ursa Major). Frank wrote several works on the ceremonies dedicated to the veneration of bears, on the relation with the constellation Ursa Major, and on the legends and tales of European populations with these animals as main characters (see for instance, Frank 1998).

The bear was considered a mediator between Gods and men, and it may be possible that he would have played such a role for tens of thousand of years. According to Lissner (1961, p. 233), who spent some time living in contact with Gilyaks (a Tungus population of the region of Amur, between Russia and China), this may explain why the bear was so important in the life of Gilyaks, and why the relationship between man and bear "played a role whose significance can scarcely be assessed today and is unknown to the majority of our contemporaries".

## 5. Bear hunting and ceremonialism

According to Hallowell's (1926, p. 40) three general types of bear hunting techniques were traditionally used from Scandinavia across northern Eurasia and North America to Labrador: a) the animal was sought in its lair and, being forced out by the hunters, was as a rule dispatched with a spear or axe as it emerged, or shot with the bow and arrow; b) the bear was frequently attacked in the open (even after it came out from its den) in what often amounted to a kind of "hand to hand" combat in which the favourite weapon was the spear or lance; c) the bear was trapped by any one of a number of devices, most frequently of the deadfall variety. According to an Aino the hunting technique was: "Drawing the knife, rushes into the animals embrace, hugs him closely and thrusts the knife home into his heart" (Hallowell 1926, p. 39). Several cave art and archaeozoology records show the hunting of bears by Palaeolithic people; for example there are Upper Palaeolithic depictions of bears that could represent hunting scenes, such as the Magdalenian bear representations of Grotte des Trois-Frères and Grotte de Massat in France (Germonpré & Hämäläinen 2007). Moreover, the idea that bears were hunted during the Middle and Upper Palaeolithic was corroborated by the discoveries in sites such as Biache-Saint-Vaast (Auguste, 2003), while no evidence exists for a possible symbolic use of bears by Neanderthals, although a worked bear incisor was found in a Chatelperronian horizon at the Grotte du Renne, Arcy-sur-Cure (David 2002).

---

[13] It may be worth to recall Frazer's (1922, p. 517) final comment in Ch. LII: "Thus the apparent contradiction in the practice of these tribes, who venerate and almost deify the animals which they habitually hunt, kill, and eat, is not so flagrant as at first sight it appears to us: the people have reasons, and some very practical reasons, for acting as they do. [… We] must endeavour to place ourselves at [an Aino's] point of view, to see things as he sees them, and to divest ourselves of the prepossessions which tinge so deeply our own views of the world. If we do so, we shall probably discover that, however absurd his conduct may appear to us, the savage nevertheless generally acts on a train of reasoning which seems to him in harmony with the facts of his limited experience".

[14] In more recent discussions the probable importance of the matrilineality has been pointed out (e.g. Allen et al., 2008), while the matriarchy has been generally excluded.

It is "more than likely that a bear cult was one of the characteristic features of an ancient Boreal culture, Old World in origin and closely associated with the pursuit of reindeer. Later, it became intercontinental in its scope, extending from Labrador to Lapland. As this culture spread… a veneration of the bear and simple rites connected with hunting the animal became more and more widely diffused and radically modified in the course of time. This hypothesis would account, it seems to me, for the ostensible differences in the customs described, as well as for the peculiar underlying trends and similarities observed" (Hallowell 1926, 161-162). This author went on to speculate that bear ceremonialism may have originated among some Palaeolithic peoples and persisted among hunting peoples of the North for millennia. If such were the case, one would expect to find some evidence of bear hunting and ceremonialism in archaeological contexts from the Late Pleistocene and Holocene. According to McLaren et al. (2005), "despite the case made for the historic-geographic hypothesis and the association of bear and cultural remains in many archaeological contexts, evidence for the practice, antiquity, and continuity of bear ceremonialism remains controversial".

Germonpré & Hämäläinen (2007) suggested that there existed a proto bear-ceremonialism in the Upper Palaeolithic, remarking the colour traces found on bear remains from Belgian caves. The red ochre traces were shown to have been applied purposely by prehistoric people and were not the result of contamination with spilt ochre or ochre containing sediment, and the archaeozoological record shows that bear elements with colour traces are mainly remains of the head and paw regions. This parallels the ethnographic evidence since in many circumpolar societies the bear head/skull and paws were colored with red or black marks during bear rituals. This bear ceremonialism dates about 26,000 BP and 23,600 BP, and the examples of manipulated bear remains in Belgium, Europe, and North America could be interpreted as with continuous bear-related rituals that started with a proto bear-ceremonialism dating from the Gravettian, and possibly even from the Aurignacian. According to Jacobson (1993) the Siberian bear cult reveals references that date from a pre-shamanic stage of prehistory, and in the Evenk mythology the bear is a culture hero from whom people received fire and stone tools, which suggests that the Evenk bear cult originated in deep antiquity.

Frazer (1922, p. 506) described the bear ceremonies of Aino of Japan[15]. A male bear cub is caught and brought into the village, where is fed for two or three years, shut up in a wooden cage. Before the killing ceremony takes place the Aino apologise to their gods, alleging that they have treated the bear kindly as long as they could, now they can feed him no longer. When the animal has been strangled to death, prayers are then addressed to him; amongst other things it is sometimes invited to return into the world in order that it may again be reared for sacrifice. During their ceremony, the Aino of Saghalien (Frazer, 1922, p. 509) say to the bear "We are holding a great festival in your honour. Be not afraid. We will not hurt you. We will only kill you and send you to the god of the forest who loves you. We are about to offer you a good dinner, the best you have ever eaten among us, and we will all weep for you together. […] You will ask God to send us, for the winter, plenty of otters and sables, and for the summer, seals and fish in abundance. Do not forget our messages, we love you much, and our children will never forget you". At the end of their prayers, they kill the bear with an arrow. The Gilyaks hold a festival of the same sort once a year. A thick layer of fat on the captive bear gives the signal for the ceremony, which is always held in winter (Frazer, 1922, p. 512). The bear is tied to a peg and shot dead with arrows. The Gilyaks look on the bear in the light of an envoy despatched with presents to the Lord of the Mountain, on whom the welfare of the people depends. Moreover, by partaking of the flesh, blood, or broth of the bear, these populations are all of opinion that they acquire some portion of the animal's mighty powers, particularly his courage and strength. No wonder, therefore, that they should treat so great a benefactor with marks of the highest respect and affection (Frazer, 1922, p. 515).

---

[15] Very similar ceremonies were performed also by other populations, such as the Sami, in the northernmost part of Eurasia (Munro 1963), and the Gilyaks (Lissner 1961).

To sum up, the indigenous populations of North Eurasia and North America shared a peculiar relationship with the bear, a sort of kinship, and adopted similar worship and ceremonialism, that could have originated in remote antiquity. Given the importance of this animal, it may be possible that even the myths regarding the Bear have an ancient origin, and in particular those that are related to his/her cosmic significance. We will recall in Sect. 7 that the constellation Ursa Major has been actually identified by northern populations not only as a female bear, but also as a cow-elk or a female mammoth with a cosmic significance.

## 6. The myth of Callisto

Ancient Greeks collected, summarizing and adding their own ideas, what was known since antiquity and was often handed down only orally. Maybe some of their myths, such as that of Callisto, the nymph that was changed first in she-bear and then in Ursa Major, could date back to the dawn of time, as we say when we usually refer to an ancient origin in prehistoric times.

Callisto, daughter of Lycaon, King of Arcadia, was one of the companions of Artemis, the virgin goddess of the hunt[16]. Even the term Arcadia may perhaps derive from *arktos*, and it is curious that such a name has indicated the fine art produced in Europe during the baroque period by artists that took inspiration from Greek myths. The Arcadia of the nymphs was anything but refined and polished; it was supposed to be more feral than wild. There were perhaps human sacrifices and (ritual) cannibalism, as one might infer from the story told by Ovid in *Metamorphoses*. The King Lycaon offered human flesh to Zeus (Jupiter) as a meal; the god transformed the King in wolf (werewolf), and then punished the horrendous human impiety with the great flood.

As it is known from Greek mythology, Zeus transformed himself in different ways in order to attract the nymphs and other beauties. In the case of Callisto he was particularly treacherous because according to some authors he took on the appearance of Artemis (Diana) in order to reach the nymph and seduce her. She then gave birth to a son, called Arcas. Various ancient writers have reported quite different versions of the myth (see for example the comments of Condos 1997, and Santoni 2009). However, all agree that since Callisto had betrayed the vow of virginity, specific of the companions of Artemis, then she was punished by the goddess and (or) by Hera (Juno), and transformed into a bear. According to Ovid's detailed tale, while the bear lived in the forests of Arcadia, the little Arcas grew with the men, and became a hunter. One day he was in a forest frequented by his mother. Callisto recognized him and went up to her son. Of course, Arcas did not think that the bear was his mother, he was frightened, and used a weapon to defend himself; in Ovid's narrative, Arcas tried to hit her in the chest[17]. At the last moment, Jupiter intervened and transformed the bear Callisto into the constellation Ursa Major, and Arcas into another constellation, maybe Arctophylax (name equivalent to guide of the bear), corresponding to the current Boötes.

---

[16] In Athens there was the Stoa of Artemis Brauronia (Brauroneion, 5th century BC), a sanctuary of Artemis on the Acropolis; and there were other shrines of Artemis in Greece, such as that of Brauron. In the Brauroneion a ritual was celebrated marking the transition of the girls to the age of marriage; during these ceremonies, the girls danced mimicking the behavior of a bear, becoming in a certain sense little bears (e.g. Lanciano and Tutino 2005).

[17] Arcas "*incidit in matrem, quae restitit Arcade viso et cognoscenti similis fuit: ille refugit inmotosque oculos in se sine fine tenentem nescius extimuit propiusque accedere aventi vulnifico fuerat fixurus pectora telo*" (Ovidius, *Metamorphoses*, 2, 500-504). It may be interesting to note that, according to Lissner (1961), during the ritual killing, the Gilyaks enraged the bear so that it was rearing, since a bear is truly vulnerable, especially to "primitive weapons", only when he rears.

Now let us consider again the Palaeolithic period. About 56,000 years ago our constellations had the look shown in Figure 4. Arcturus means bear guard, and it seems to us that the name gets a sense when it is referred not to the sky of the Greeks, but to that of tens of thousand of years earlier. That time Ursa Major included a fifth star, i.e. it had a pentagonal shape, and also the Ursa Minor had a very different shape. One could then ask if they were such shapes that inspired the "core" of the myth, i.e. the meeting between the mother bear and the man (Figure 5), a myth that was then handed down in some way during the following millennia.

Of course, our interpretation could be just imagination. Moreover, we must point out several obvious caveats. Not only there are rather different versions of the myth of Callisto, but there are also many different interpretations about what the Ursa Major stars should represent, apart from a Big Dipper. The Mesopotamians called them Big Wagon or Chariot, the Latins Seven oxen (*Septem triones*), while for the Egyptians they were the thigh of a bull. Another problem is that there is no incontrovertible proof of the absence of cultural contamination between Eurasia and America after the thaw and before the discovery of 1492. The question is not just about a European culture that possibly crossed the Atlantic; for example, Rasmussen et al. (2010) have suggested another migration from Siberia to North America and Greenland, which occurred approximately 5500 years ago, long after the thaw. Finally, one cannot explain scientifically the ancient myths: for thousands of years they were handed down orally, modified, adapted, and translated in languages unknown to us. However, despite these limitations, our exercise shows that, with the scientific methods, one can perhaps guess something of the temporal stratification of ancient myths, and grasp a possible meaning of the core of the myth of Callisto.

Our approach doesn't look much different from that mentioned for example by Anisimov. The ethnographic "monuments" are only the vestiges of "primitive life", since they are far removed in time from the present-day ethnographically surviving forms. In "these forms, modified over time, we can discover to some extent the basic ties of their historical relationship. By following the traces of the sources of the phenomenon being studied, we can reconstruct scientifically the process of development in its various phases" (Anisimov 1963, p. 159).

Let us consider for instance a possible "stratum" of the myth. When Hera realized that the nymph-bear had been transformed into a constellation, she got angry and asked the Ocean, the great river that surrounded the ancient inhabited land, not to allow the Bear to bathe in its waters. In other words, this part of the myth would explain why the constellation is always above the horizon, i.e. it is circumpolar. This was the case at the time of the ancient Greeks, at the latitude of Athens, and even today several stars of the constellation are circumpolar. The precession, over many thousands of years, provides a different picture, so this part of the myth should not be as old as the celestial encounter between the mother Bear and the Hunter suggested by us. The rotation axis of the Earth moves, albeit very slowly, and as a result the North Pole, which today is close to the Polar Star $\alpha$ Ursae Minoris, at a distance of thousands of years was (and will be) in another location. The period of the precession cycle is of about 26,000 years. In 2000 BC the North Pole was located in between the two Bears[18], and it had been in a similar position also in the millennia centred around 28,000 BC and 54,000 BC. Note, however, that the cycle is not regular, therefore the similarity is only approximated and should be intended within some degrees. In 16000 BC the celestial pole was located in the constellation Cygnus, and that time Ursa Major, seen from a mean latitude location, was not circumpolar, that is, due to the sky rotation all its stars could go below the horizon[19].

---

[18] According to Porphyry, the Pithagoreans considered Ursa Major and Ursa Minor the two hands of the goddess Rhea (see e.g. Zhmud 2012, note 117 p. 199). This was recalled by DeSantillana and Von Dechend (1977, p. 3) in relation to their hypothesis of the rotating cosmic mill; it may have a better sense if we take into account the ancient way of considering the sky as seen from outside.

[19] It may be worth to remark that also in the millennia centred around 42,000 BC the North Pole was located in Cygnus, and the two Bears were farther from the polar region. At mean latitudes, only Ursa Minor was (partly) circumpolar; Ursa Major was visible mainly during the nights of spring and summer, approximately in a position corresponding to the present day Hercules.

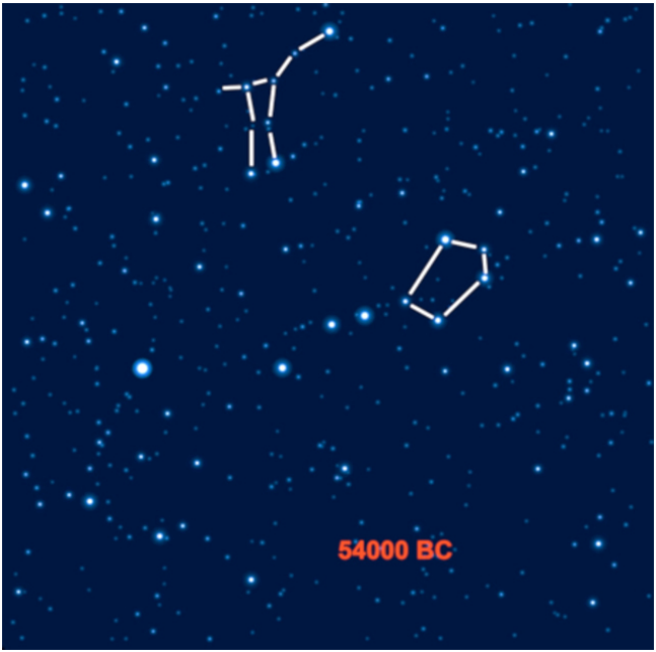

**Fig. 4.** About 56,000 years ago the shapes of the constellations Ursa Minor and Ursa Major were quite different from the present one.

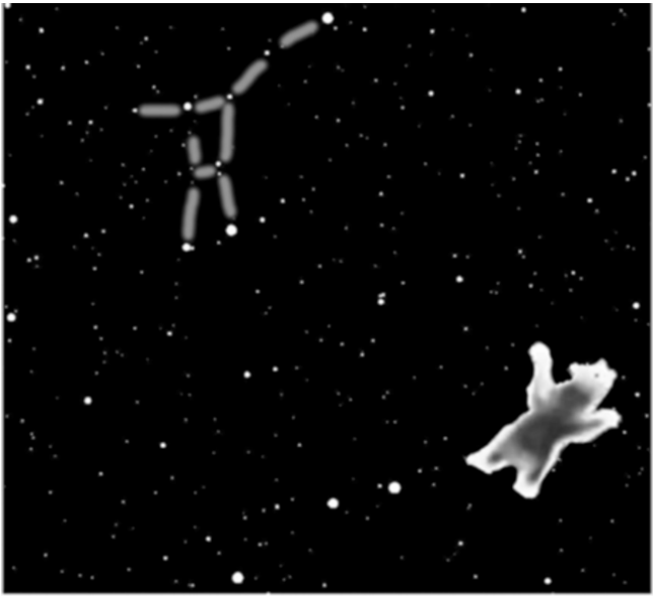

**Fig. 5.** A possible interpretation of the constellations 56,000 years ago.

**7. Other cosmic myths**

According to the legends of some native American tribes, the quadrangle of Ursa Major[20] represents a bear, while the other three bright stars are hunters in its pursuit. The celestial hunting takes place throughout the year, from when the bear comes out of hibernation in the spring, until the killing in autumn[21]. However, while these tribes and the ancient Greek and Latin writers speak of a bear, some Scandinavian and Nordic populations have interpreted the constellation as elk (moose), with horns though a female, and it seems there are some Siberian native legends telling instead about a female mammoth. There are many variations of the stories among the northern peoples of America and Eurasia[22]. In most cases these myths deal with a cosmic hunt, and Vasilevich (1963) and Anisimov (1963) discussed for example those of Siberian populations. Some tales include also Ursa Minor as a cub of the adult animal represented by Ursa Major.

---

[20] In classical and European tradition, instead, the constellation Ursa Major is larger, and includes about 20 visible stars.
[21] See e.g. the book for children written by Monroe and Williamson (1987).

In Siberia, almost all Evenks call the constellation of the Great Bear *kheglun*, and it means elk. In particular the Evenks Bachin (Tunguska) told a legend very similar to that of the three North American hunters that were chasing the Great Bear; in this case the four stars of the quadrangle were not a bear but an elk. A variation among Vitim-Olekma Evenks told of a giant that hunted the heavenly elk on skis and cut off its head and tail; the rest of the elk, the body and legs, was changed into a constellation, while the ski track was changed into the Milky Way. For other tribes of Evenks, it was a cosmic bear, Mangi, with a dual nature, half man and half animal (mythical figure of the first ancestor), who hunted the elk, and he was represented in the sky by Boötes and Arcturus. According to the Evenks Urmi, the master-spirit of the upper world hunted a great elk cow and her fawn. The elk got away to the sea, but she took a bone from the hind leg of her fawn and brought it to Earth. In this way, elks appeared on the earth. "The mother who disappeared became kheli (the mammoth [sea lion?]) and remained in the sea" (Vasilevich 1963; Anisimov, p. 162, concluded that the cosmic image of the elk is understood as that of a "mother-elk").

"The linking of the cosmological image of the elk (deer) with the image of the mammoth, functionally analogous, and, in turn, that of mammoth with the views on the nether world, is characteristic of the mythology of most of the nationalities of northern Asia and it may be considered a common Siberian phenomenon" (Anisimov, 1963). Maybe it is interesting to note that the extinct mammoth was related to the nether world. Ethnographic materials show that the link of the elk-maral image with the Sun is one of the most ancient elements of the cosmological concepts of the majority of peoples of Siberia. The male elk, associated with the Sun, is visible during the day in the celestial taiga, while at night it is the female elk with her calf to be visible. Clearly, the cosmic hunt should have reflected the forms of economic activities of "primitive man" (Anisimov, 1963, p. 163).

Finally, it may be worth to recall the mistress (or mistresses) of the universe, mother of animals or more generally life-giver of the universe, simultaneously imaged as both a woman and an animal; for example the Eskimo Sedna is conceived both as a woman and a walrus-cow. According to Anismov (1963, p. 169), the image of mother-animal at first was not applied to all animals, but only to species of totem-animal. "This means that concepts of the so-called mythical first ancestors, constituting the center of the entire mythology of the totemists, in all probability must have borne the image of a female spirit originally: the mythical ancestress-animal".

**8. Conclusion**

According to paleoanthropologists, one of the main activities of our ancestors of the Upper Palaeolithic of Eurasia and North America was the hunt, as it has been for many indigenous populations up to modern times. It is remarkable that most of the myths and legends concerning the constellation of the Big Dipper, collected by ethnographers among the indigenous populations of northern latitudes, have to do with the hunt of a female animal (bear, elk, mammoth), and this animal has a cosmic importance. Maybe there is some relation with the possible primeval matrilineality.

It may be possible that even the core of the myth of Callisto regarding Ursa Major, the Great she-Bear, is very ancient, and that in the Upper Palaeolithic there was a bear ceremonialism similar to that of northern indigenous populations described by the ethnographers. Of course, the antiquity of the Great she-Bear cannot be demonstrated. We have tried to show, however, that the reconstruction of the sky of the past millennia could help in grasping something of the relation of the ancient man with the sky, supplying complementary information to those obtained (or deduced) from archaeological and ethnographical records as regards the life and the possible mentality of the man of Palaeolithic.

---

[22] See e.g. Y. Berezkin, *The cosmic hunt: variants of a siberian-north-american myth*, and the references therein, in particular the Russian authors. http://www.folklore.ee/folklore/vol31/berezkin.pdf.